\documentclass[preprint,review,12pt]{elsarticle}

\usepackage{amsmath}
\usepackage{amssymb}

\usepackage{graphicx}% Include figure files
\usepackage{bm}% bold math
\renewcommand{\vec}[1]{\boldsymbol{#1}}

\def\cpc{{\itshape Comput. Phys. Commun.} }
\def\pop{{\itshape Phys. Plasmas} }
\def\jcp{{\itshape J. Comput. Phys.} }
\def\jgr{{\itshape J. Geophys. Res.} }
\def\apj{{\itshape Astrophys. J.} }
\def\ppcf{{\itshape Plasma Phys. Control. Fusion} }

\journal{Computer Physics Communications}

\begin{document}

\begin{frontmatter}

\title{Hyper Boris integrators for kinetic plasma simulations}

\author[a,b]{Seiji Zenitani\corref{author}}
\author[c]{Tsunehiko N. Kato}
\cortext[author] {Corresponding author.\\\textit{E-mail address:} seiji.zenitani@oeaw.ac.at}
\address[a]{Space Research Institute, Austrian Academy of Sciences, Schmiedlstra{\ss}e 6, 8042 Graz, Austria}
\address[b]{Research Center for Urban Safety and Security, Kobe University, 1-1 Rokkodai-cho, Nada-ku, Kobe 657-8501, Japan}
\address[c]{Graduate School of Artificial Intelligence and Science, Rikkyo University, Tokyo 171-8501, Japan}

\begin{abstract}
We propose a family of numerical solvers for
the nonrelativistic Newton--Lorentz equation in kinetic plasma simulations.
The new solvers extend the standard 4-step Boris procedure,
which has second-order accuracy in time, in three ways. 
First, we repeat the 4-step procedure multiple times,
using an $n$-times smaller timestep ($\Delta t/n$).
We derive a formula for the arbitrary subcycling number $n$,
so that we obtain the result without repeating the same calculations.
Second, prior to the 4-step procedure,
we apply Boris-type gyrophase corrections to the electromagnetic field.
In addition to a well-known correction to the magnetic field,
we correct the electric field in an anisotropic manner
to achieve higher-order ($N=2,4,6 \dots$th order) accuracy.
Third, combining these two methods,
we propose a family of high-accuracy particle solvers,
{\itshape the hyper Boris solvers},
which have two hyperparameters of the subcycling number $n$ and
the order of accuracy, $N$.
The $n$-cycle $N$th-order solver gives
a numerical error of $\sim (\Delta t/n)^{N}$
at affordable computational cost.
\end{abstract}


\end{frontmatter}

\section{Introduction}
Kinetic plasma simulations such as 
particle-in-cell (PIC) simulation \citep{hockney,birdsall,v05}
and hybrid simulation \citep{lipatov02}
are very useful for understanding complex phenomena
in space, solar, and astrophysical plasmas. 
These simulations resolve Lagrange motions of many charged particles,
typically $10^2$--$10^3$ particles in a grid cell, and
$10^{8}$--$10^{12}$ particles in the entire simulation domain. 
Even though they are computationally very expensive,
kinetic simulations are capable of reproducing various kinetic processes
from the first principle.

One of the most important components of kinetic plasma simulations is
a particle integrator, also referred to as a particle pusher,
which advances individual particles in the electromagnetic field
by using the Newton--Lorentz equation. 
%Since the integrator is used for all the particles, 
%its accuracy, stability, and computational cost is crucial for the simulations.
One of the most popular particle integrators is the Boris solver \citep{boris70},
or the so-called Buneman--Boris solver.
The Boris solver is relatively simple,
consists of four-step operations as will be shown in this paper,
and it is proven to preserve the phase-space volume \citep{qin13}.
The solver has a second-order accuracy in time,
i.e., it produces an error $\propto (\Delta t)^2$ in velocity,
where $\Delta t$ is the timestep. 
Owing to its simplicity, numerical stability, and good accuracy,
it has been used as the de facto standard method for over a half century. 
%In fact, together with the current-deposition schemes,
%the particle solver is one of the two most expensive parts in modern PIC simulation.

Since there is always a demand for better simulation results,
scientists have been actively developing high-accuracy integrators
\citep{patacchini09,he15,winkel15,umeda18,umeda19,zeni18,zeni20,kato21}.
In addition, modern processors are so fast that
the data transfer inside the chip or from the memory
often becomes a bottleneck for simulations. 
In such a case, even though computationally expensive,
higher-order or higher-accuracy integrators
that provide good results without extra data transfer
are favorable.
Nevertheless, most of the current integrators have second-order accuracy,
and they do not drastically outperform the second-order Boris solver.
The number of higher-order particle integrators is still limited \citep{winkel15}. 

In this study, we propose
a new family of high-accuracy Boris-type integrators
for nonrelativistic kinetic simulations.
Three enhancements to the standard Boris solver are presented.
In Section \ref{sec:review}, we briefly review basic issues.
We outline the standard Boris solver and
its gyrophase correction in Section \ref{sec:boris}, and
the multiple Boris solver \citep{zeni20}
which subcycles the Lorentz-force part of the Newton--Lorentz equation
in Section \ref{sec:multiboris}.
In Section \ref{sec:multicycle},
we propose a new subcycling method that repeats the entire 4-step procedure.
In Section \ref{sec:higher},
we propose higher-order corrections to the 4-step Boris solver.
In Section \ref{sec:hyper}, combining the two methods,
we propose a family of high-accuracy particle solvers,
{\itshape the hyper Boris solvers}.
It has two hyperparameters of the subcycling number $n$ and
the order of accuracy, $N$.
Section \ref{sec:test} presents numerical tests of the proposed methods.
Section \ref{sec:discussion} contains discussion and summary.

\section{Review}
\label{sec:review}

\subsection{Boris solver}
\label{sec:boris}

Here we discuss a popular implementation of the Boris solver \citep{boris70}.
We limit our attention to the nonrelativistic particle motion.
We use SI units. 
%\begin{align}
%\frac{d\vec{x}}{d t}
%&= \vec{v} \\
%\frac{d\vec{v}}{d t}
%&= 
%\frac{q}{m} \left( \vec{E} + {\vec{v}} \times \vec{B} \right)
%\label{eq:acc_phys}
%\end{align}
We consider the equation of motion in a leap-frog and discrete manner,
\begin{align}
\frac{\vec{x}^{t+\Delta t/2} - \vec{x}^{t-\Delta t/2}}{\Delta t}
&= \vec{v}^{t} \\
\frac{\vec{v}^{t+\Delta t} - \vec{v}^{t}}{\Delta t}
&= 
\frac{q}{m} \left( \vec{E}^{t+\frac{\Delta t}{2}} + \frac{{\vec{v}^{t+\Delta t} + \vec{v}^{t}}}{2}\times \vec{B}^{t+\frac{\Delta t}{2}} \right)
\label{eq:acc}
\end{align}
%where symbols have their standard meanings.
where $\vec{x}$ is the position, $\vec{v}$ is the velocity,
$q$ is the charge, $m$ is the mass of the particle,
$\vec{E}$ is the electric field at the particle position,
$\vec{B}$ is the magnetic field, and
$\Delta t$ is the timestep.
The superscript indicates the time, i.e.,
$\vec{v}^t$ is the velocity of the particle at time $t$.
%We assume that
%the electromagnetic fields are stationary
The electromagnetic fields are customarily assumed uniform and stationary
from $t$ and $t+\Delta t$, and
we drop the superscripts from $\vec{E}$ and $\vec{B}$ for simplicity.
We further define the element magnetic/electric vectors,
\begin{align}
\vec{\tau}_1 &\equiv \frac{q\Delta t}{2m} \vec{B},
~~~
\vec{\varepsilon}_1 \equiv \frac{q\Delta t}{2m} \vec{E}
\end{align}
These vectors are basic elements of numerical procedures in this paper. 
Eq.~\eqref{eq:acc} approximates
an exact solution for the next velocity $\vec{v}^{t+\Delta t}$,
\begin{align}
\vec{v}^{t+\Delta t}
&=
\cos(2\tau_1) \vec{v}^{t}
+
2 {\rm sinc}(2\tau_1) (\vec{v}^{t} \times \vec{\tau}_1+ \vec{\varepsilon}_1 )
\nonumber \\ 
&~~~
+
2 {\rm sinc}^2(\tau_1)
\Big(
(\vec{v}^{t}  \cdot \vec{\tau}_1 ) \vec{\tau}_1 
+
\vec{\varepsilon}_1 \times \vec{\tau}_1
\Big)
%\nonumber \\ 
%&~~~
+
\frac{2(1-{\rm sinc}(2\tau_1))}{\tau_1^2}
(\vec{\varepsilon} \cdot \vec{\tau}_1 ) \vec{\tau}_1 
,
\label{eq:analytic}
\end{align}
where ${\rm sinc}(x)\equiv \sin(x)/x$ is the sinc function
and $\tau_1= |\vec{\tau}_1|$.
For reference, Eq.~\eqref{eq:analytic} is derived in \ref{sec:appendix}.
When $\tau_1=0$,
Eq.~\eqref{eq:analytic} contains divisions by zero
inside the sinc function and in the coefficient of the last term,
and therefore the equation is not convenient for numerical calculation.

The Boris solver gives a second-order approximation of Eq.~\eqref{eq:analytic}.
We tentatively set $\vec{t}_1 = \vec{\tau}_1$ and $\vec{e}_1 = \vec{\varepsilon}_1$.
It accelerates the particle velocity by the following 4 steps.
\begin{eqnarray}
\left\{
\begin{array}{ll}
\vec{v}^{-} &= \vec{v}^{t} + \vec{e}_1
\label{eq:boris1} \\
\vec{v}^{'} &= 
\vec{v}^{-} + \vec{v}^{-}\times \vec{t}_1
\label{eq:boris2} \\
\vec{v}^{+} &= \vec{v}^{-} + \dfrac{2}{ 1 + t_1^2 }~\vec{v}^{'}\times \vec{t}_1
\label{eq:boris3}
\\
\vec{v}^{t+\Delta t} &= \vec{v}^{+} + \vec{e}_1
\label{eq:boris4}
\end{array}
\right.
\label{eq:4step}
\end{eqnarray}
Here, $\vec{v}^{+}, \vec{v}^{-}, \vec{v}'$ are intermediate velocities.
The first and fourth equations represent the half-acceleration by $\vec{E}$, and
the second and third equations stand for the gyration about $\vec{B}$.
The gyration part is schematically illustrated by
the gray dashed lines in Fig. 1(a).
One can see that the gyration, the circular motion in the velocity space,
is approximated by a combination of two triangles.
The phase angle $\theta $ is approximated by
\begin{align}
\theta = \dfrac{qB}{m} \Delta t = 2\tau_1 \approx 2 \alpha_1 \equiv 2 \arctan \tau_1
\label{eq:theta}
\end{align}
where $\alpha_1 = \arctan \tau_1$ is a half angle of the approximated gyration.
Examining the tangent vector from $\vec{v}^{-}$ to $\vec{v}'$,
one can see the ratio between the true tangent to the approximation is
\begin{align}
f(\tau_1) = \frac{ \tan \tau_1 }{ \tau_1 } = 1 + \frac{1}{3} \tau_1^2 + \frac{2}{15} \tau_1^4  + \frac{17}{315} \tau_1^6 + \dots
\label{eq:gc_taylor}
\end{align}
This indicates that there is a second-order error in the tangent line
and in other places accordingly.
The second-order error is usually small, and therefore
the 4-step procedure (Eq.~\eqref{eq:4step}) is widely used.
We call the 4-step procedure the standard Boris solver
(This corresponds to the Boris-B solver in \citet{zeni18}).

To cancel this error (Eq.~\eqref{eq:gc_taylor}),
\citet{boris70} amplified the element magnetic vector
by a factor of $f_{N}=f_{N}({\tau}_1)$,
taken from the Taylor expansion up to the $(N-2)$th order ($N=2,4,6,\dots$).
For example, for $N=6$,
\begin{align}
\vec{t}_1
\equiv f_{6}(\tau_1)\vec{\tau_1}
= \left( 1 + \frac{1}{3} \tau_1^2 + \frac{2}{15} \tau_1^4 \right) \vec{\tau_1}
\label{eq:gyrophase_6th_B}
\end{align}
Using this $\vec{t}_1$, we use the standard 4-step procedure.
Then we can improve the gyration part to the $N$th-order accuracy: 
$2\alpha_{\rm 1,corr} = \theta + \mathcal{O}(\tau_1^N)$,
where the subscript ${\rm corr}$ indicates the corrected value.
The corrected gyration is illustrated by
the gray solid lines in Fig. 1(a).
This correction is known as the gyrophase correction.
Note that we only amplifies the magnetic vector $\vec{\tau_1}$ for advancing an individual particle.
We do not amplify the magnetic field $\vec{B}^{t+\frac{\Delta t}{2}}$ to calculate the fields at the next timestep $\vec{B}^{t+\frac{3\Delta t}{2}}$ and $\vec{E}^{t+\frac{3\Delta t}{2}}$.
We call the entire procedure
the Boris solver with gyrophase correction
(the Boris-A solver in Ref.~\citep{zeni18}).

The gyrophase correction allows us to
solve the gyration part to a higher-order of accuracy; however,
the solver consists of the gyration part and
the half-acceleration parts by the electric field. 
Unless $\vec{E}=0$,
the entire 4-step procedure falls back to a second-order accuracy \citep{zeni18}.
For this reason, the standard Boris solver
without the gyrophase correction is popularly used.

\begin{figure*}[htbp]
\begin{center}
\includegraphics[width={0.85\textwidth}]{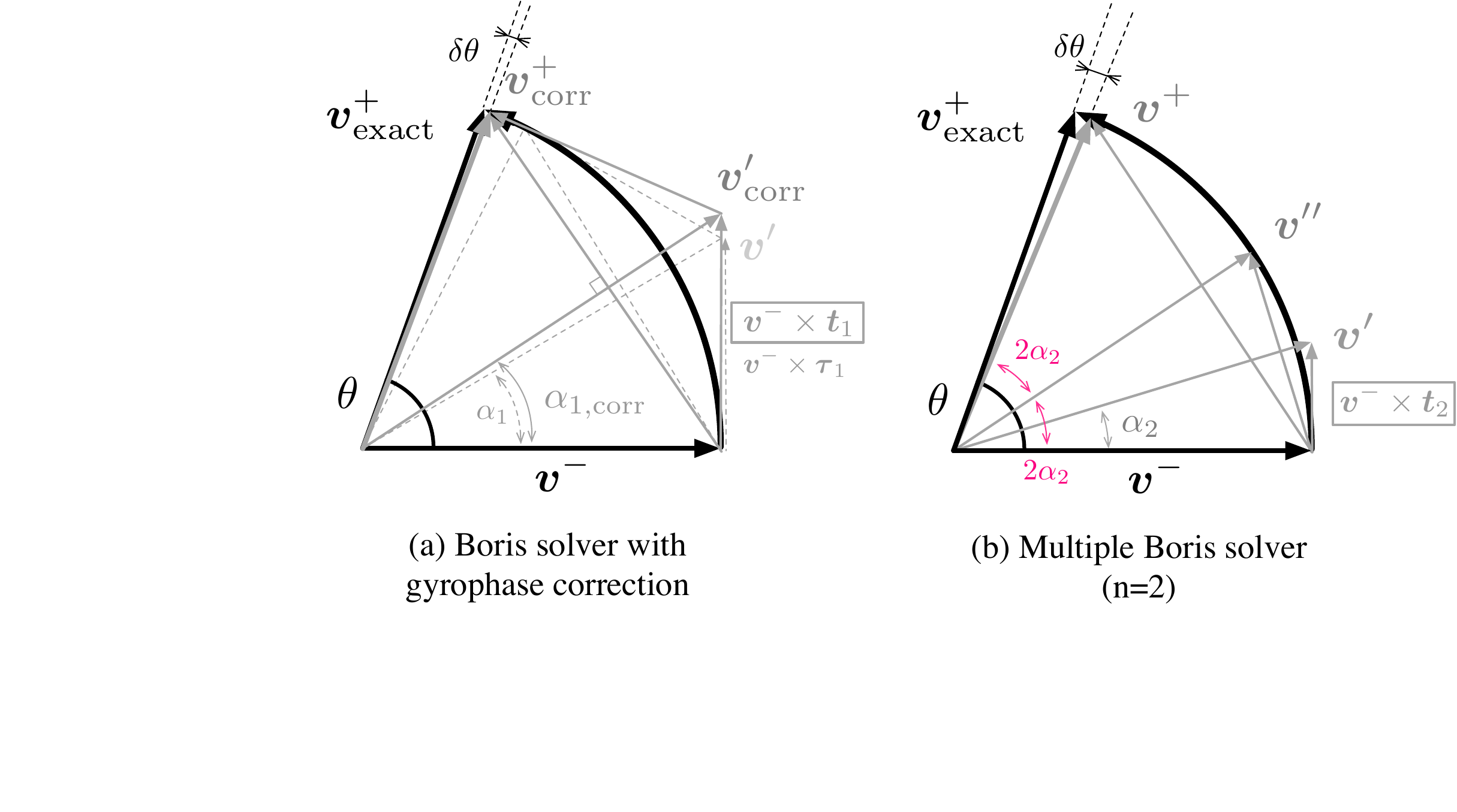}
\caption{
(a) Schematic diagram of
the Lorentz-force part of 
the Boris solver with gyrophase correction.
The subscript `corr' emphasizes a corrected value.
(b) The Lorentz-force part of
the multiple Boris solver \cite{zeni20} with $n=2$.
\label{fig:scheme}}
\end{center}
\end{figure*}

\subsection{Multiple Boris solver}
\label{sec:multiboris}

Next, we outline the multiple Boris solver \citep{zeni20},
which subcycles the gyration part of the Boris solver.
The middle two procedures in Eq.~\eqref{eq:4step}
that advance the pre-gyration state $\vec{v}^-$ to the post-gyration state $\vec{v}^+$
are repeated $n$ times with a $n$-times smaller timestep, $\Delta t/n$.
The gyration part of the $n=2$ case is illustrated in Figure \ref{fig:scheme}(b).
Here, we use the element magnetic vector $\vec{t}_n$ and the relevant half angle $\alpha_n$,
\begin{align}
\vec{t}_n \equiv \vec{\tau}_n \equiv \frac{\vec{\tau}_1}{n} %= \frac{q\Delta t}{2nm}\vec{B}
,~~
\alpha_n \equiv \arctan t_n
.\label{eq:multiBoris_defs}
\end{align}
\citet{zeni20} have derived
the post-gyration state $\vec{v}^+$ of the multiple Boris solver
for an arbitrary positive integer $n$.
\begin{align}
\vec{v}^+
&= 
c_{n1} \vec{v}^-
+
c_{n2} ~ (\vec{v}^- \times \vec{t}_n)
+
c_{n3}
( \vec{v}^- \cdot \vec{t}_n ) \vec{t}_n
\label{eq:multiBoris}
\end{align}
where $c_{n1}$, $c_{n2}$, and $c_{n3}$ are the following coefficients,
\begin{align}
c_{n1}
&=
\cos(2n\alpha_n) = 
T_{n} (p_n)
\label{eq:cn1}
\\
c_{n2}
&=
\cfrac{\sin(2n\alpha_n)}{t_n} =
\frac{2}{1+t_n^2}
U_{n-1} (p_n)
\label{eq:cn2}
\\
c_{n3}
&= \cfrac{ 1-\cos(2n\alpha_n) }{t_n^2} 
=
\left\{
\begin{array}{ll}
\dfrac{2}{1+t_n^2}
& {\rm (for~{\it n}=1)}
\\
\dfrac{2}{1+t_n^2}
\Big(
U_{k}(p_n)
+
U_{k-1}(p_n)
\Big)^2
& {\rm (for~{\it n}=2k+1)}
\\
\dfrac{8}{(1+t_n^2)^2}
\Big(
U_{k-1}
\big(p_n)
\Big)^2
& {\rm (for~{\it n}=2k)}
\end{array}
\right.
\label{eq:cn3}
\end{align}
where $T_n(x)$ and $U_n(x)$ are Chebyshev polynomials of the first kind and the second kind,
$k$ in Eq.~\eqref{eq:cn3} is a positive integer,
and $p_n$ is the cosine parameter,
\begin{align}
p_n \equiv \cos 2\alpha_n = \frac{1-t_n^2}{1+t_n^2}
.
\end{align}
Importantly, by using Eq.~\eqref{eq:multiBoris},
we can immediately obtain the post-gyration state $\vec{v}^+$
without repeating the 2-step calculation $n$ times.
The numerical error of the gyration part is $\propto (\Delta t/n)^2$,
as expected \citep{zeni20}.

\section{Multicycle solver}
\label{sec:multicycle}

Motivated by the multiple Boris solver,
we propose to subcycle the entire 4-step procedure of the particle integrator. 
%Using an arbitrary integer $n$, we consider
%a small timestep $\Delta t \rightarrow \Delta t/n$, and then
%we repeat the 4-step procedure (Eq.~\eqref{eq:4step}) $n$ times.
%This technique is called subcycling, and occasionally used to improve the particle motion \cite{adam82,arefiev15,hirvijoki20}.
We call this method the multicycle method.
Note that the multiple Boris solver subcycles the Lorentz-force part inside the 4-step procedure, but the multicycle solver subcycles the entire 4-step procedure.

We first define
\begin{align}
\vec{t}_n \equiv \vec{\tau}_n \equiv \frac{\vec{\tau}_1}{n} %= \frac{q\Delta t}{2nm}\vec{B}
,~~
\vec{e}_n \equiv\vec{\varepsilon}_n \equiv \frac{\vec{\varepsilon}_1}{n} %= \frac{q\Delta t}{2nm}\vec{E}
,~~
\alpha_n \equiv \arctan t_n,
\label{eq:n_vectors}
\end{align}
We rewrite $\vec{v}^{(0)} = \vec{v}^{t}$, $\vec{v}^{(n)} = \vec{v}^{t+\Delta t}$,
and the intermediate states $\vec{v}^{(1)}, \vec{v}^{(2)}, \cdots, \vec{v}^{(n-1)}$.
Each subcycle advances $\vec{v}^{(k)}$ to $\vec{v}^{(k+1)}$.
The entire procedure is equivalent to
\begin{eqnarray}
\left.
\begin{array}{c}
\left\{
\begin{array}{ll}
\vec{v}^{-} &= \vec{v}^{(0)} + \vec{e}_n
\label{eq:bb1} \\
\vec{v}^{+} &= \mathbb{R}_{\vec{b}}(2\alpha_n) ~\vec{v}^{-}
\label{eq:bb2} \\
\vec{v}^{(1)} &= \vec{v}^{+} + \vec{e}_n
\label{eq:bb3}
\end{array}
\right.
\\
\vdots
\\
\left\{
\begin{array}{ll}
\vec{v}^{-} &= \vec{v}^{(n-1)} + \vec{e}_n
\label{eq:bb4} \\
\vec{v}^{+} &= \mathbb{R}_{\vec{b}}(2\alpha_n) ~\vec{v}^{-}
\label{eq:bb5} \\
\vec{v}^{(n)} &= \vec{v}^{+} + \vec{e}_n
\label{eq:bb6}
\end{array}
\right.
\end{array}
~~~~~~~
\right\}
\times ~n~{\rm cycles}
\label{eq:ntimes0}
\end{eqnarray}
Here we express the 4-step procedure in three lines.
The gyration part is rewritten by a rotation operator $\mathbb{R}_{\vec{b}}(2\alpha_n)$ on the pre-gyration state $\vec{v}^{-}$.
It stands for a rotation about the $\vec{b} \equiv \vec{B}/|B|$ axis
with a rotation angle $2\alpha_n$.

After $n$ cycles, the result of Eq.~\eqref{eq:ntimes0} is given by
\begin{align}
\vec{v}^{(n)}
&=
\mathbb{R}_{\vec{b}}(2n\alpha_n) ~\vec{v}^{(0)}
+
\bigg( \mathbb{R}_{\vec{b}}(2n\alpha_n) + 2 \sum_{j=1}^{n-1} \mathbb{R}_{\vec{b}}(2j\alpha_n) + \mathbb{I} \bigg) ~\vec{e}_n
\nonumber\\
&=
\mathbb{R}_{\vec{b}}(2n\alpha_n) ~\vec{v}^{(0)}
+
\sum_{j=0}^{n-1} \bigg( \mathbb{R}_{\vec{b}}(2(j+1)\alpha_n) + \mathbb{R}_{\vec{b}}(2j\alpha_n) \bigg) ~\vec{e}_n
.
\label{eq:vn}
\end{align}
In the multiple Boris solver (Sec.~\ref{sec:multiboris}), we notice that
the post-gyration vector $\vec{v}^{+}$ is rotated from $\vec{v}^{-}$ with an angle $2\alpha_n$.
Then we can apply the multi-Boris formula (Eq.~\eqref{eq:multiBoris})
to the first term of the right hand side of Eq.~\eqref{eq:vn},
\begin{align}
\mathbb{R}_{\vec{b}}(2n\alpha_n) ~\vec{v}^{(0)}
&=
c_{n1} \vec{v}^{(0)}
+
c_{n2} ~ (\vec{v}^{(0)} \times \vec{t}_n)
+
c_{n3}
( \vec{v}^{(0)} \cdot \vec{t}_n ) \vec{t}_n
\label{eq:cycle_eq1}
\end{align}
The coefficients $c_{n1}, c_{n2}, c_{n3}$ are given by Eqs.~\eqref{eq:cn1}--\eqref{eq:cn3}. 
%The $\alpha_n$ approximates a half angle of the single Boris rotation.
We similarly apply the formula to
the second term of the right hand side of Eq.~\eqref{eq:vn}, and then obtain
\begin{align}
&
\sum_{j=0}^{n-1} \bigg( \mathbb{R}_{\vec{b}}(2(j+1)\alpha_n) + \mathbb{R}_{\vec{b}}(2j\alpha_n) \bigg) ~\vec{e}_n
~~~~~~~~~~~~~~~~~~~~~~~~~
\nonumber \\
&~~~~
=
\Big( \sum_{j=0}^{n-1} \left( c_{j+1,1} + c_{j1} \right) \Big)
\vec{e}_n
+
\Big( \sum_{j=0}^{n-1} \left( c_{j+1,2} + c_{j2} \right) \Big)
(\vec{e}_n \times \vec{t}_n)
+
\Big( \sum_{j=0}^{n-1} \left( c_{j+1,3} + c_{j3} \right) \Big)
( \vec{e}_n \cdot \vec{t}_n ) \vec{t}_n
,
\label{eq:cycle_eq2}
\end{align}
where $c_{n1}, c_{n2}, c_{n3}$ are the coefficients by Eqs.~\eqref{eq:cn1}--\eqref{eq:cn3}.
%Then, we set the three coefficients in the right hand side
%to $c_{n4}$, $c_{n5}$, and $c_{n6}$. 
%\begin{align}
%\label{eq:cn456}
%c_{n4} \equiv \sum_{k=0}^{n-1} \left( c_{k+1,1} + c_{k1} \right),~~
%c_{n5} \equiv \sum_{k=0}^{n-1} \left( c_{k+1,2} + c_{k2} \right),~~
%c_{n6} \equiv \sum_{k=0}^{n-1} \left( c_{k+1,3} + c_{k3} \right).
%\end{align}
We further examine the terms in the right hand side of Eq.~\eqref{eq:cycle_eq2}.
Using Eqs.~\eqref{eq:cn1} and \eqref{eq:cn2}, we find
\begin{align}
\sum_{j=0}^{n-1} \left( c_{j+1,1} + c_{j1} \right)
&= 
\frac{\sin\alpha_n}{\sin\alpha_n}\cdot 
\sum_{j=0}^{n-1}
\Big( \cos(2j+2)\alpha_n + \cos2j\alpha_n \Big)
\nonumber\\
&= 
\frac{\cos\alpha_n}{\sin\alpha_n}\cdot 
\sum_{j=0}^{n-1}
\Big(
\sin(2j+2)\alpha_n - \sin 2j\alpha_n
\Big) %\cos \alpha_n
\nonumber\\
&
%= \frac{\sin 2n\alpha_n}{\tan \alpha_n}
= \frac{\sin 2n\alpha_n}{t_n} = c_{n2}
\label{eq:cn4}
.
\end{align}
Using Eqs.~\eqref{eq:cn1}--\eqref{eq:cn3}, we similarly find
\begin{align}
\sum_{j=0}^{n-1} \left( c_{j+1,2} + c_{j2} \right)
&=
\frac{\sin\alpha_n}{\sin\alpha_n} \cdot 
\frac{1}{t_n}
\sum_{j=0}^{n-1}   \Big( \sin (2j+2)\alpha_n  + \sin 2j\alpha_n \Big)
\nonumber\\
&=
-
\frac{\cos\alpha_n}{\sin\alpha_n} \cdot 
\frac{1}{t_n}
\sum_{j=0}^{n-1}
\Big( \cos(2j+2)\alpha_n  -
\cos 2j\alpha_n  \Big)
\nonumber\\
&= \frac{1 - \cos 2n\alpha_n}{t^2_n} = c_{n3}
.
\label{eq:cn5}
\end{align}
Using Eqs.~\eqref{eq:cn1}--\eqref{eq:cn3} and \eqref{eq:cn4}, we obtain
\begin{align}
\sum_{j=0}^{n-1} \left( c_{j+1,3} + c_{j3} \right)
&=
\frac{1}{t_n^2}
\left(
2n
%- \sum_{j=0}^{n-1}
%Big(\frac{ \cos(2j+2)\alpha_n + \cos 2j\alpha_n }{t_n^2}\Big)
- 
\sum_{j=0}^{n-1} \left( c_{j+1,1} + c_{j1} \right)
\right)
= \frac{2n - c_{n2}}{t_n^2}
%= \frac{2}{t_n^2} \bigg( n - \frac{1}{1+t_n^2} U_{n-1}\Big(\frac{1-t_n^2}{1+t_n^2}\Big) \bigg)
= \frac{2}{t_n^2} \bigg( n - \frac{1}{1+t_n^2} U_{n-1}(p_n) \bigg)
\label{eq:cn6}
\end{align}

Summarizing Eqs.~\eqref{eq:vn}--\eqref{eq:cn6} and
recovering the original notation of $\vec{v}^{t+\Delta t}$ and $\vec{v}^{t}$,
we obtain the following form
\begin{align}
\vec{v}^{t+\Delta t}
&=
c_{n1} \vec{v}^{t}
+
c_{n2} ~ (\vec{v}^{t} \times \vec{t}_n)
+
c_{n3}
( \vec{v}^{t} \cdot \vec{t}_n ) \vec{t}_n
%\nonumber \\ 
%&~~~
+
c_{n4} \vec{e}_n
+
c_{n5} (\vec{e}_n \times \vec{t}_n)
+
c_{n6}
( \vec{e}_n \cdot \vec{t}_n ) \vec{t}_n
\label{eq:multicycle}
\end{align}
where we use the following coefficients, $c_{n1} \dots c_{n6}$.
The first three coefficients and the cosine parameter $p_n$ are identical
to ones in Section \ref{sec:multiboris}, but we show them here for completeness. 
\begin{align}
c_{n1}
&=
T_{n} \left(p_n\right)
\label{eq:coeff01}
\\
c_{n2}
&=
c_{n4}
=
\frac{2}{1+t_n^2}
U_{n-1} \left(p_n\right)
\label{eq:coeff02}
\\
c_{n3}
&=
c_{n5}
%\nonumber
%\\
%&=
=
\left\{
\begin{array}{ll}
\dfrac{2}{1+t_n^2}
& {\rm (for~{\it n}=1)}
\\
\dfrac{2}{1+t_n^2}
\left(
U_{k}\left(p_n\right)
+
U_{k-1}\left(p_n\right)
\right)^2
& {\rm (for~{\it n}=2k+1)}
\\
\dfrac{8}{(1+t_n^2)^2}
\left(
U_{k-1}
\left(p_n\right)
\right)^2
& {\rm (for~{\it n}=2k)}
\end{array}
\right.
\label{eq:coeff03}
\\
c_{n6}
&
= \frac{2}{t_n^2} \left( n - \frac{1}{1+t_n^2} U_{n-1}\left(p_n\right) \right)
%= (1-p_n)\Big( \frac{2n}{1+p_n}- U_{n-1}(p_n) \Big)
\label{eq:coeff06}
\\
p_n &= \frac{1-t_n^2}{1+t_n^2}
\end{align}
The multicycle formula (Eq.~\eqref{eq:multicycle}) holds true, even for $\vec{t}_n=0$.
In practice, we can just use Eq.~\eqref{eq:multicycle} to obtain $\vec{v}^{t+\Delta t}$.
This is much quicker than repeating the 4-step procedure $n$ times.

The new solver will be tested in Section \ref{sec:test}.
For reference, we list some coefficients here ($n=1,2,4$).
\begin{align}
c_{11} = \frac{1-t_1^2}{1+t_1^2},~~
c_{12} = c_{13} = c_{14} = c_{15} = c_{16} = \frac{2}{1+t_1^2},
\label{eq:c1}
\end{align}
\begin{align}
c_{21} = \frac{1-6t_2^2+t_2^4}{(1+t_2^2)^2},~~
c_{22} = c_{24} = \frac{4(1-t_2^2)}{(1+t_2^2)^2},~~
c_{23} = c_{25} = \frac{8}{(1+t_2^2)^2},~~
c_{26} =
\frac{4(3+t_2^2)}{(1+t_2^2)^2},
\label{eq:c2}
\end{align}
\begin{align}
c_{41}
= \frac{1-28t_4^2+70t_4^4-28t_4^6+t_4^8}{(1+t_4^2)^4},~~
%\nonumber\\
% c_{42} = \frac{8(1-t^2)(1-6t^2+t^4)}{(1+t^2)^4},~~
c_{42} = c_{44} = \frac{8(1-7t_4^2+7t_4^4-t_4^6)}{(1+t_4^2)^4},~~
\nonumber \\
c_{43} = c_{45} = \frac{32(1-t_4^2)^2}{(1+t_4^2)^4},~~
c_{46} = \frac{8(11-t_4^2+5t_4^4+t_4^6)}{(1+t_4^2)^4}
\label{eq:c4}
\end{align}

\section{Higher-order correction}
\label{sec:higher}

In this section, we develop a higher-order correction.
We consider the standard Boris solver without gyrophase correction.
We retain $\vec{t}_1 \equiv \vec{\tau}_1$ and $\vec{e}_1 \equiv\vec{\varepsilon}_1$.
Substituting $n=1$ into Eq.~\eqref{eq:multicycle},
we rewrite the 4-step procedure (Eq.~\eqref{eq:4step})
\begin{align}
\vec{v}^{t+\Delta t}
&=
\frac{1-t_1^2}{1+t_1^2}
\vec{v}^t
+
\frac{2}{1+t_1^2}
\vec{v}^t \times \vec{t}_1
+
\frac{2}{1+t_1^2}
\left( \vec{v}^t \cdot \vec{t}_1 \right) \vec{t}_1
\nonumber\\
&~~~+
\frac{2}{1+t_1^2} \vec{e}_1
+
\frac{2}{1+t_1^2}
\vec{e}_1\times \vec{t}_1
+
\frac{2}{1+t_1^2}
( \vec{e}_1 \cdot \vec{t}_1 ) \vec{t}_1
\label{eq:boris_et}
\end{align}
For a moment we assume $|\vec{t}_1|\ne 0$.
We rewrite the fourth term of the right hand side, by using a vector triplet
$-(\vec{e}_1\times\vec{t}_1)\times\vec{t}_1
= t_1^2\vec{e}_1 - {(\vec{e}_1\cdot\vec{t}_1)\vec{t}_1}$,
\begin{align}
\vec{v}^{t+\Delta t}
&=
\frac{1-t_1^2}{1+t_1^2}
\vec{v}^t
+
\frac{2}{1+t_1^2}
\vec{v}^t \times \vec{t}_1
+
\frac{2}{1+t_1^2}
\left( \vec{v}^t \cdot \vec{t}_1 \right) \vec{t}_1
\nonumber \\
&~~~+
\frac{2}{1+t_1^2}
\vec{e}_1\times \vec{t}_1
-
\frac{2}{1+t_1^2}
\frac{(\vec{e}_1\times\vec{t}_1)}{t_1^2}\times\vec{t}_1
+
2 \frac{( \vec{e}_1 \cdot \vec{t}_1 )}{t_1^2}\vec{t}_1
\end{align}
then we further arrange the terms.
%We further define a unit vector $\vec{\hat{b}}\equiv \vec{t}_1/t_1=\vec{B}/B$.
%and then obtain
\begin{align}
\left( 
\vec{v}^{t+\Delta t}
-\frac{\vec{e}_1\times\vec{t}_1}{t_1^2}
\right)
&=
\frac{1-t_1^2}{1+t_1^2}
\left(
\vec{v}^t - \frac{\vec{e}_1\times\vec{t}_1}{t_1^2}
\right)
+
\frac{2}{1+t_1^2}
\left(
\vec{v}^t - \frac{\vec{e}_1\times\vec{t}_1}{t_1^2}
\right)
\times \vec{t}_1
%\frac{2t_1}{1+t_1^2}
%\left(
%\vec{v}^t - \frac{\vec{e}_1\times\vec{t}_1}{t_1^2}
%\right)
%\times \vec{\hat{b}}
\nonumber\\
&~~~
+
\frac{2}{1+t_1^2}
\left( \vec{v}^t \cdot \vec{t}_1 \right) \vec{t}_1
%\frac{2t_1^2}{1+t_1^2}
%( \vec{v}^t \cdot \vec{\hat{b}} ) \vec{\hat{b}}
+
%2 (\vec{e}_1 \cdot \vec{\hat{b}} )\vec{\hat{b}}
2 \frac{( \vec{e}_1 \cdot \vec{t}_1 )\vec{t}_1}{t_1^2}
\label{eq:phys0}
\end{align}
From Eqs.~\eqref{eq:cn1} and \eqref{eq:cn2},
we find $\sin 2\alpha_1 = 2t_1/(1+t_1^2)$ and $\cos 2\alpha_1 = (1-t_1^2)/(1+t_1^2)$.
Then we rewrite Eq.~\eqref{eq:phys0} in the physics notation,
\begin{align}
\left( \vec{v}^{t+\Delta t} - \frac{\vec{E} \times \vec{B}}{B^2} \right)
 &= \left( \vec{v}^t - \frac{\vec{E} \times \vec{B}}{B^2} \right) \cos 2\alpha_1
+ \left( \Big( \vec{v}^t - \frac{\vec{E} \times \vec{B}}{B^2} \Big) \times \vec{\hat{b}} \right) \sin 2\alpha_1
\nonumber \\
&~~~
+ \left( \vec{v}^t - \frac{\vec{E} \times \vec{B}}{B^2} \right)_{\parallel} (1-\cos 2\alpha_1) + \frac{q\vec{E}_{\parallel}}{m}\Delta t
\label{eq:phys}
\end{align}
where $\vec{\hat{b}}\equiv \vec{t}_1/t_1=\vec{B}/B$ is the unit vector parallel to the magnetic field and
the subscript $\parallel$ indicates the parallel component of the vectors.
This is identical to the time reversible formula,
earlier derived by \citet{buneman67} (Eq.~(50) in Ref.~\citep{buneman67}), except the last term,
because Ref.~\citep{buneman67} assumed $\vec{E}_\parallel = 0$.
Comparing Eq.~\eqref{eq:phys} and Rodrigues' rotation formula,
we immediately see that Eq.~\eqref{eq:phys} indicates
gyration around the {\bf E} $\times$ {\bf B} velocity with the angle of $2\alpha_1$.
In fact, Eq.~\eqref{eq:phys} is
a good approximation of the exact solution of the particle velocity
in a constant electromagnetic field:
\begin{align}
\left( \vec{v}^{t+\Delta t}
 - \frac{\vec{E} \times \vec{B}}{B^2} \right)
 &= \left( \vec{v}^t - \frac{\vec{E} \times \vec{B}}{B^2} \right) \cos 2\tau_1  + \bigg( \Big( \vec{v}^t - \frac{\vec{E} \times \vec{B}}{B^2} \Big) \times \vec{\hat{b}} \bigg) \sin 2\tau_1
\nonumber\\&~~~~
+ \left( \vec{v}^t - \frac{\vec{E} \times \vec{B}}{B^2} \right)_{\parallel} (1-\cos 2\tau_1 ) + \frac{q\vec{E}_{\parallel}}{m}\Delta t
\label{eq:exact}
\end{align}
The two equations (\eqref{eq:phys} and \eqref{eq:exact}) only differ
in the rotation angle, $2\tau_1 \ne 2\alpha_1$.

Using these equations,
we discuss the influence of the gyrophase correction, presented in Section \ref{sec:boris}.
Following Eqs.~\eqref{eq:gc_taylor} and \eqref{eq:gyrophase_6th_B}, we amplify
the element magnetic vector $\vec{t}_1 = f_N(\tau_1)\vec{\tau}_1$
by a factor $f_N(\tau_1)$ ($f_N$ for short in this section; $N=2,4,6,\cdots$).
Then Eq.~\eqref{eq:phys} can be rewritten to
\begin{align}
\left( \vec{v}^{t+\Delta t}
 - \frac{\vec{E} \times \vec{B}}{f_N B^2} \right)
 &= \left( \vec{v}^t - \frac{\vec{E} \times \vec{B}}{f_N B^2} \right) \cos 2\alpha_{1,corr}
+ \left( \Big( \vec{v}^t - \frac{\vec{E} \times \vec{B}}{f_N B^2} \Big) \times \vec{\hat{b}} \right) \sin 2\alpha_{1,corr}
\nonumber \\
&~~~
+ \left( \vec{v}^t - \frac{\vec{E} \times \vec{B}}{f_N B^2} \right)_{\parallel} (1-\cos 2\alpha_{1,corr}) + \frac{q\vec{E}_{\parallel}}{m}\Delta t
\label{eq:phys2}
\end{align}
Note that the magnetic field is amplified to $\vec{B} \rightarrow f_N \vec{B}$ and
that the gyration angle is corrected to $2\alpha_1 \rightarrow 2\alpha_{1,corr}$. 
Eq.~\eqref{eq:phys2} tells us that the gyrophase correction makes
the {\bf E}$\times${\bf B} velocity unphysically slow,
\begin{align}
\frac{\vec{E} \times \vec{B}}{B^2}
\rightarrow
\frac{\vec{E} \times \vec{B}}{f_N B^2}
%= \frac{\vec{E} \times \vec{B}}{B^2}\left(1-\frac{1}{3}\tau_1^2+\mathcal{O}(\tau_1^4)\right)
\label{eq:etau1}
\end{align}
In other words, this is the reason why the Boris solver with higher-order gyrophase correction
fall back to the second-order accuracy.

To deal with this, we propose to amplify
the electric field perpendicular to the magnetic field $\vec{E}_{\perp}$
by a factor of $f_N$: $\vec{E}_{\perp} \rightarrow f_N\vec{E}_{\perp}$.
Then we can eliminate the unwanted $f_N$ in Eq.~\eqref{eq:etau1}.
Importantly, we amplify only the perpendicular components of the element electric vector,
but retain the parallel component,
in order not to modify the last term in Eq.~\eqref{eq:phys2}. 
With this in mind, we amplify the element electric vector in the following way,
\begin{align}
\vec{e}_1
\equiv f_N \vec{\varepsilon}_1 + \Big( 1-f_N \Big)
\frac{( \vec{\varepsilon}_1 \cdot \vec{\tau}_1 ) \vec{\tau}_1}{\tau_1^2} \label{eq:gyrophase_6th_E}
\end{align}
%This gives $f_N\vec{\varepsilon}_{1\perp}$ in the perpendicular direction and $\vec{\varepsilon}_{1\parallel}$ in the parallel direction.
For example, for $N=6$, we have
\begin{align}
\vec{e}_1
\equiv \left(1+\frac{1}{3}\tau_1^2+\frac{2}{15}\tau_1^4\right) \vec{\varepsilon}_1 - \left(\frac{1}{3}+\frac{2}{15}\tau_1^2\right) ( \vec{\varepsilon}_1 \cdot \vec{\tau}_1 ) \vec{\tau}_1
.
\label{eq:gyrophase_6th_E2}
\end{align}
Then we use $\vec{e}_1$ in the Boris's 4-step procedure.
These equations work for $\vec{\tau}_1=0$, too.
For example, if we substitute $\vec{\tau}_1=0$ into Eq.~\eqref{eq:gyrophase_6th_E2}, we just obtain $\vec{e}_1
= \vec{\epsilon}_1$.

In summary, we propose to correct both
the magnetic vector $\vec{t}_1$ (Eq.~\eqref{eq:gyrophase_6th_B}) and
the electric vector $\vec{e}_1$ (Eq.~\eqref{eq:gyrophase_6th_E}),
before using the 4-step procedure.
Using the $(N-2)$th-order Taylor polynomial ($N=2,4,6,\dots$) of Eq.~\eqref{eq:gc_taylor},
the two prescriptions (Eqs.~\eqref{eq:gyrophase_6th_B} and \eqref{eq:gyrophase_6th_E})
can be easily extended. 
Then the 4-step procedure provides
a solution with an $N$th-order accuracy,
as will be tested in Section \ref{sec:test}.

\begin{figure}[htbp]
\begin{center}
\includegraphics[width={0.8\textwidth}]{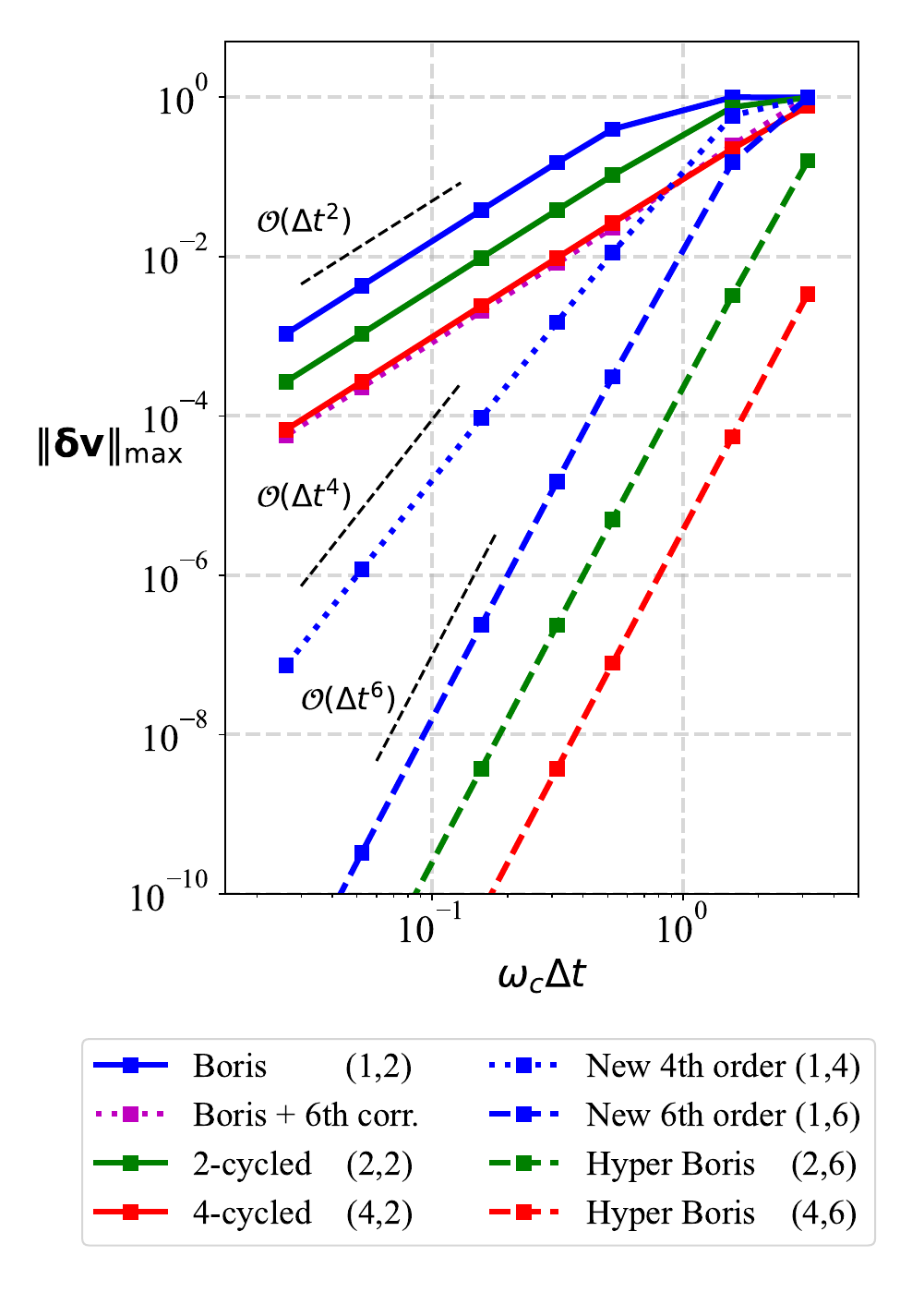}
\caption{
Scaling of maximum numerical errors in the velocity $\vec{v}$,
as a function of the timestep $\omega_c \Delta t$.
The label $(n,N)$ indicates $n$-cycled $N$th order solver.
\label{fig:err}}
\end{center}
\end{figure}

\section{Hyper Boris solver}
\label{sec:hyper}

Both the multicycle method (Section \ref{sec:multicycle}) and
the higher-order correction (Section \ref{sec:higher})
use the standard 4-step procedure (Eq.~\eqref{eq:4step}).
The former repeats the 4-step procedure multiple times, and
the latter corrects the electromagnetic vectors prior to the 4-step procedure.
In other words, the two methods do not modify the 4-step procedure itself.
Recognizing this fact, we propose a hybrid solver of
the multicycle method and the higher-order correction
to achieve even better accuracy.
In practice, we corrects the electromagnetic vectors,
before using the multicycle procedure.
The entire procedure is as follows.
\begin{enumerate}
\item
(Advance the particle position ${\vec{x}^{t+\Delta t/2} \leftarrow \vec{x}^{t-\Delta t/2}} + \vec{v}^{t} {\Delta t}$)
\item
Define the element magnetic and electric vectors, $\vec{\tau}_n$ and $\vec{\varepsilon}_n$ (Eq.~\eqref{eq:n_vectors})
\item
Referring to Eq.~\eqref{eq:gc_taylor},
set $f_N(\tau_n)$ to a truncated Taylor expansion of $f(\tau_n)$ up to $(N-2)$th order.
Note that this is not $\tau_1$ but $\tau_n$. 
\item
Referring to Eqs.~\eqref{eq:gyrophase_6th_B} and \eqref{eq:gyrophase_6th_E},
calculate the magnetic and electric vectors $\vec{t}_n$ and $\vec{e}_n$
\begin{align}
\vec{t}_n
\equiv f_N\,\vec{\tau_n},
~~~
\vec{e}_n
\equiv f_N\,\vec{\varepsilon}_n + \Big( 1-f_N \Big)
\frac{( \vec{\varepsilon}_n \cdot \vec{\tau}_n ) \vec{\tau}_n}{\tau_n^2} 
\end{align}
\item
Calculate the four coefficients $c_{n1},c_{n2},c_{n3}$, and $c_{n6}$ (Eqs.~\eqref{eq:coeff01}--\eqref{eq:coeff06})
\item
Compute the next state $\vec{v}^{t+\Delta t}$ by the multicycle formula (Eq.~\eqref{eq:multicycle})
\end{enumerate}
%The multicycle method and the higher-order correction correspond to the $(n,2)$ and $(1,N)$ cases of the hyper Boris solvers.
We call this {\itshape the hyper Boris solver},
because it has two hyperparameters, $n$ and $N$.
One can consider arbitrary combinations of
the subcycling number $n$ and $N$th-order accuracy. 
Their combination excellently works,
as will be shown in Section \ref{sec:test}.
%Since one can combine because it is the $n$-cycle and the  solvers.

\section{Numerical tests}
\label{sec:test}

In order to test the proposed solvers,
we have carried out test-particle simulations
in a static electromagnetic field.
The following solvers are compared:
\begin{enumerate}
\item
Boris solver with and without the 6th-order gyrophase correction \citep{boris70}
\item
Multicycle Boris solver ($n=2, 4$),
proposed in Section \ref{sec:multicycle}
\item
New higher-order Boris solver ($N=4,6$) in Section \ref{sec:higher}
\item
Hyper Boris solver $(n,N) = (2,6), (4,6)$ 
proposed in Section \ref{sec:hyper}
\end{enumerate}
The particle parameters are set to be $m=1$, $q=1$,
$\vec{x}=(0,0,0)$, and $\vec{v}=(0,0,0)$.
The electromagnetic fields are set to be $\vec{E}=(0,0.5,0.1)$ and $\vec{B}=(0,0,1)$. The gyro frequency is $\omega_c \equiv (qB/m) = 1$.
In each run, we evaluate an error in the velocity,
$\|\vec{v}-\vec{v}_{\rm exact}\|$, where $\vec{v}_{\rm exact}$ is given by Eq.~\eqref{eq:analytic}.
Then we have recorded the maximum error
during the six gyroperiods, $0< \omega_c t\le 12 \pi$.
The timestep ranges from $\omega_c \Delta t = \pi/120$ to $\pi$.

Figure~\ref{fig:err} shows the results as a function of the timestep $\Delta t$.
The standard Boris solver is indicated by the blue solid line.
It is evident that it provides the second-order error.
Although partially hidden by the red solid line (the $n=4$ multicycle solver),
the magenta dotted line indicates
the Boris solver with the 6th-order gyrophase correction.
It provides smaller errors than the standard Boris solver,
but eventually returns the second-order error. 
The multicycle Boris solvers, indicated by the color solid lines,
give expected results. 
They provide $n^2$ times smaller errors than the Boris solver.
By definition, the Boris solver does not give a good approximation
for $\omega_c\Delta t \gtrsim \pi/2$, but
the multicycle solvers relax this threshold to $\omega_c\Delta t \gtrsim n\pi/2$. 
The blue dotted and dashed lines indicate
the new higher-order Boris solvers.
They have $N$th-order accuracy, $\propto (\Delta t)^N$,
in contrast to the Boris solver with 6th-order gyrophase correction.
Finally, the hyper Boris solvers are indicated by the color dashed lines.
Since the two improvements work together,
they give even better results.
Now the error is proportional to $\propto (\Delta t/n)^N$.

\begin{figure}[htbp]
\begin{center}
\includegraphics[width={\columnwidth}]{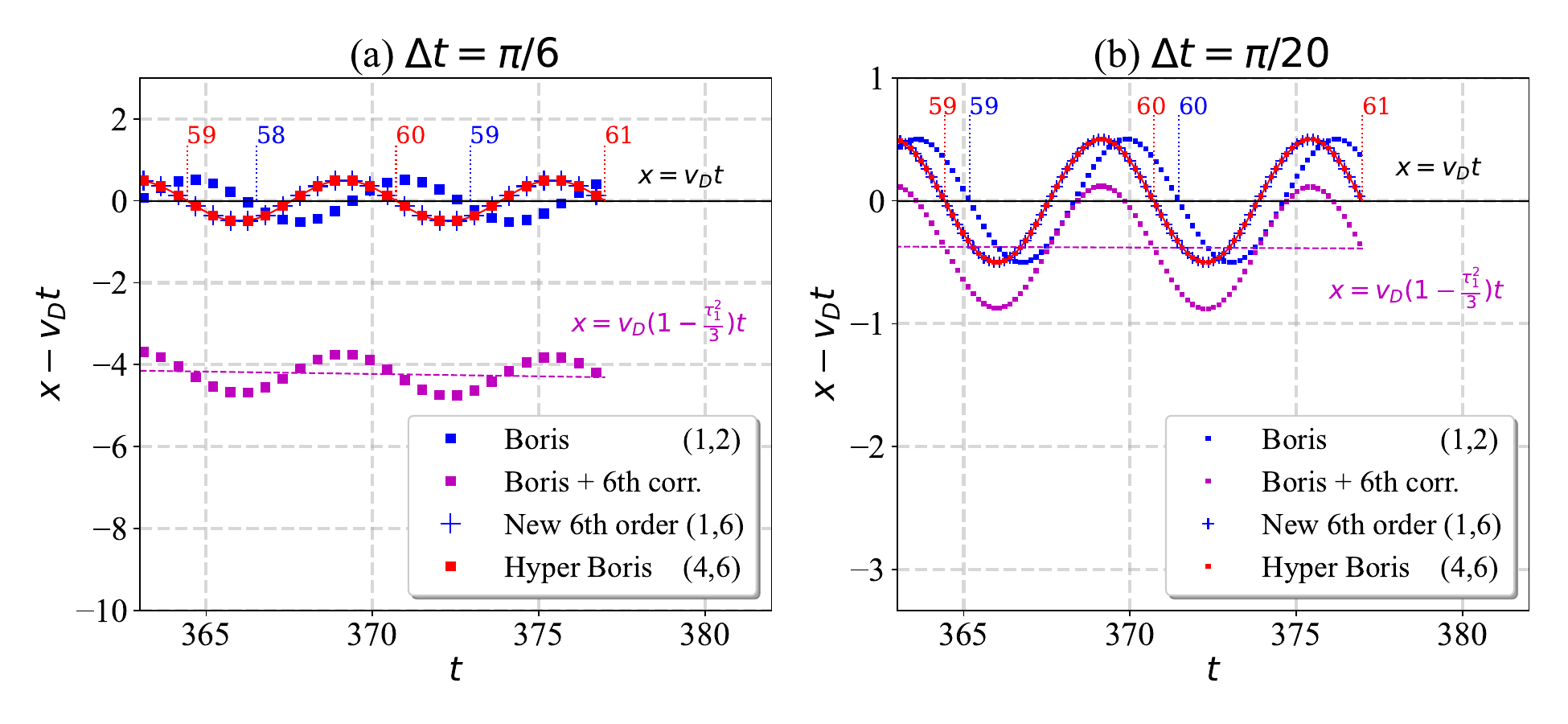}
\caption{
Particle positions in the {\bf E}$\times${\bf B} drift frame $(x-v_Dt)$
as a function of time $t$,
for (a) $\omega_c\Delta t=\pi/6$ and (b) $\omega_c\Delta t=\pi/20$. 
The drift speed is set to be $v_{D}=\|\vec{E}\times\vec{B}\|/B^2=0.5$.
Four particle solvers are compared.
The solid red curve indicates the analytic solution.
The magenta dashed line indicates
the erroneous drift effect of $\approx v_D (1-\tau_1^2/3)t$
(Eq.~\eqref{eq:errdrift}).
The red and blue numbers such as `60' indicate
the starting points of the 60th gyroperiod.
%$x$-position of particles as a function of time $t$.
%Four particle solvers are compared.
%The red line indicates the analytic solution.
%The black dashed and magenta dotted lines indicate the average drift speed,
%$v_{D}=\|\vec{E}\times\vec{B}\|/B^2$, and
%the erroneous drift speed, $v_D (1-\tau_1^2/3)$.
\label{fig:drift}}
\end{center}
\end{figure}

Next, in order to check long-term behaviors,
we run test particle simulations in the same electromagnetic field over sixty gyroperiods, $\omega_c t \le 120 \pi$.
Panels in Fig.~\ref{fig:drift} show particle positions as a function of time
for (a) $\omega_c\Delta t=\pi/6 \approx 0.52$ and (b) $\omega_c\Delta t=\pi/20 \approx 0.16$.
In these cases, while gyrating, the particles drift in the $+x$ direction.
Then the $x$-positions in the {\bf E}$\times${\bf B} drift frame are presented,
i.e., $(x-v_Dt)$, 
where $v_D=\|\vec{E}\times\vec{B}\|/B^2=0.5$ is the drift speed.
The symbols indicate the results of the four particle solvers. 
%This is indicated by the black dashed lines in Fig.~\ref{fig:drift}. 
The red solid line indicates the analytic solution of particle position. 
One can see that the gyrophase in the standard Boris solver (the blue squares) differs from the others, because of a numerical delay in phase.
There is a larger delay in the $\omega_c\Delta t=\pi/6$ case,
because of the second-order error in gyrophase.
In contrast, even though it better reproduces the gyrophase,
the Boris solver with the 6th-order gyrophase correction (the purple squares)
moves in $+x$ at the erroneously slow speed.
This is in agreement with our estimate (Eq.~\eqref{eq:etau1}),
\begin{align}
\frac{1}{f_6}
\frac{\| \vec{E} \times \vec{B} \|}{ B^2}
=
v_D \left(1-\frac{1}{3}\tau_1^2+\mathcal{O}(\tau_1^4)\right)
\label{eq:errdrift}
\end{align}
The new 6th-order Boris solver and the $(4,6)$ hyper Boris solver
do not exhibit this drift-speed problem.
Their results are virtually indistinguishable from
the exact solutions in Figs.~\ref{fig:drift}(a) and (b). 
Strictly speaking,
there remain small displacements between
the particle positions and the exact positions,
$\|\vec{r}-\vec{r}_{\rm exact}\|=0.006 \sim 0.017 \ll r_L$,
where $r_L=|\vec{v}_{\rm t=0}-\vec{v}_D|/\omega_c=0.5$ is the typical gyroradius,
because we advance the particle by the leap-frog method.
This second-order displacement does not grow in time.

\begin{figure}[htbp]
\begin{center}
\includegraphics[width={\columnwidth}]{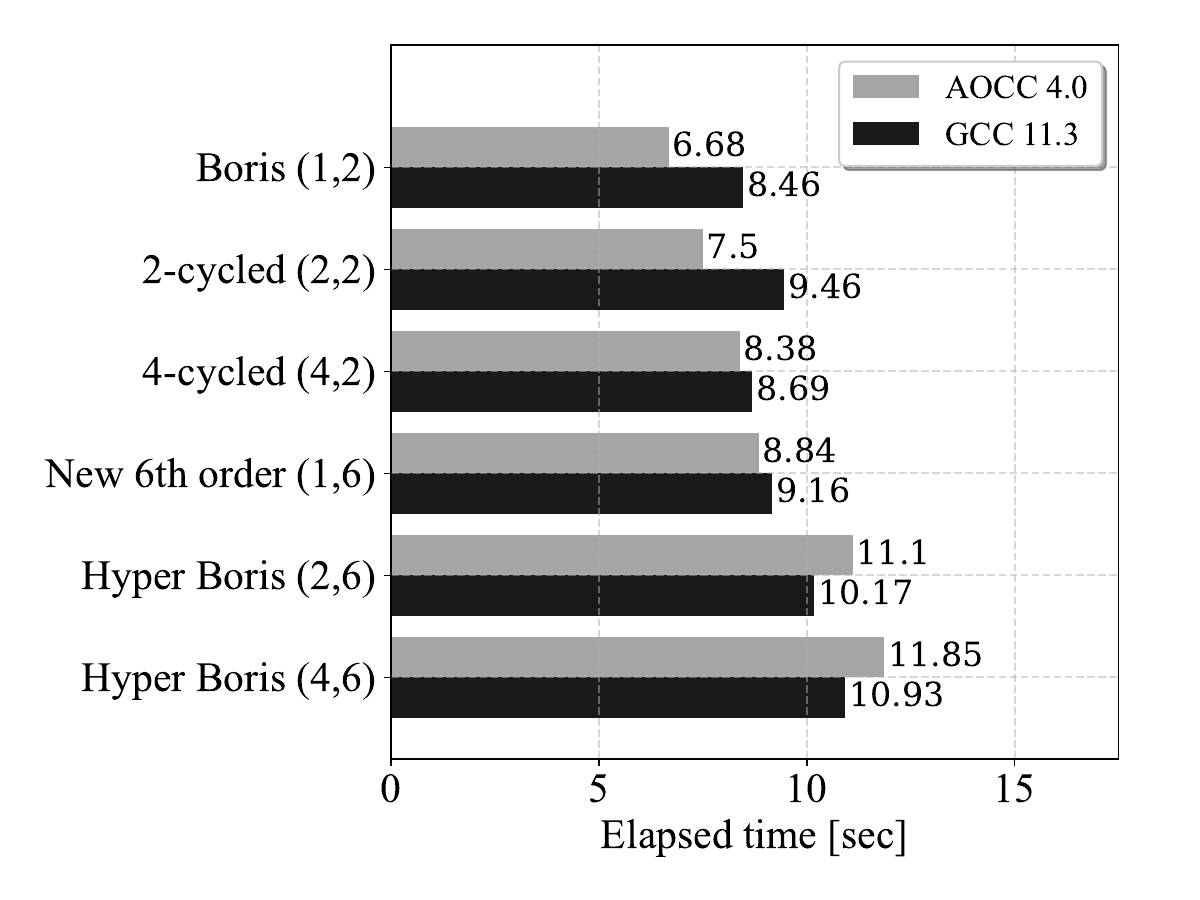}
\caption{
Elapsed time of test-particle simulations
by the Boris solver and the hyper Boris solvers.
The label $(n,N)$ indicates $n$-cycled $N$th order solver.
\label{fig:bar}}
\end{center}
\end{figure}

In order to evaluate the computational cost,
we run the same simulations with $\omega_c \Delta t=\pi/6$ until $\omega_c t=10^8 \pi$
on our Linux PC with AMD Ryzen 5955WX processor.
We use the AMD optimizing fortran compiler (AOCC) 4.0
and GNU Compiler Collection (GCC) 11.3 with the {\verb|-O2|} option.
We use the standard 4-step procedure (Eq.~\eqref{eq:4step}) for $n=1$ and
the multicycle formula (Eq.~\eqref{eq:multicycle}) for $n=2,4$.
Their elapsed times are compared in Fig.~\ref{fig:bar}.
Thanks to the formula,
the program does not need to repeat the 4-step procedure,
and therefore the numerical cost increases only reasonably.
The field correction requires additional cost.
The 4-cycled 6th-order solver is 30--75\% more expensive than the standard Boris solver.
Even though the benchmark results highly depend on
the specific implementation, compiler, and computer,
these results suggest that the proposed solvers are promising. 
For comparison, we have also implemented and tested a direct solver,
which uses Eq.~\eqref{eq:analytic} in the simulation.
In this case, as mentioned earlier,
we need to deal with the division by zero
inside the sinc function and in the last term,
because $\tau_1$ can be zero.
We only found that the direct solver is substantially slower than other integrators:
it took 25.63 (AOCC) and 19.91 (GCC) seconds in the same test.
This confirms that the approximate integrators are useful.

\section{Discussion}
\label{sec:discussion}

In this article, we have discussed
three improvements to the Boris integrator for kinetic plasma simulations.
First, we have proposed the multicycle method.
Partial or full sub-stepping of the particle integrator was proposed by many authors
\citep{adam82,friedman91,arefiev15,umeda18,umeda19,zeni20};
however, our unique point is that we have derived and utilized
the multicycle formula (Eq.~\eqref{eq:multicycle}) for arbitrary $n$
for the acceleration part of the particle integrator.
The formula contains six coefficients.
%Conveniently, we only need to calculate four,
%because the other two are identical to the others. 
Conveniently, we only need to calculate four coefficients,
because among the six coefficients there are two pairs of coefficients each of which are identical.
Second, prior to the 4-step procedure,
we use the gyrophase correction to $\vec{B}$ and
the anisotropic correction to $\vec{E}$
to achieve higher-order accuracy.
A similar idea was discussed in earlier literature
(Section 4.3 (p.60) in \citet{birdsall}; Section 4.7.1 (p.114) in \citet{hockney}), but
it has not been used over many decades. 
Third, combining the subcycling and the higher-order correction,
we have proposed the hyper Boris solver.
We summarize all the procedures in Section \ref{sec:hyper}.
It only gives a numerical error of $\propto (\Delta t/n)^{N}$,
as confirmed by numerical tests in Section \ref{sec:test}.

It is noteworthy that all the solvers are reversible.
As evident in Eq.~\eqref{eq:ntimes0},
the multicycle procedures are symmetric in time.
The higher-order correction is time-symmetric as well,
because it uses the electromagnetic fields at the time center $t+\Delta t/2$. 
So is the hyper Boris solver, because it is a hybrid of the two. 
We expect that such a time-symmetric solver is stable,
because the particle motion is unlikely to be damped or amplified in one specific direction. 
Another favorable property for the long-term stability is
the incompressibility in the phase-space volume,
often denoted as the volume-preservation \citep{qin13}.
Eq.~\eqref{eq:phys} and subsequent discussion tell us that
the Boris solver with higher-order correction gives
a rotation by the angle of $2\alpha_{1,corr}$ and
a parallel displacement by $({q\vec{E}_{\parallel}}/m)\Delta t$ in the velocity space.
These operations conserve a volume in the velocity-space, and
the hyper Boris solver inherits this property. 
Together with the position part,
the hyper Boris solver is volume preserving,
similar to the original Boris solver. 
%In addition, as already discussed,
%the hyper Boris solver is free from the zero division problem when $\vec{B}=0$. 

The hyper Boris solvers give accurate results at affordable costs.
Our numerical tests show that
the 4-cycle 6th-order solver needs 30--75\% more computational time
than the standard Boris solver.
Considering its ultrahigh accuracy, this is acceptable. 
It is also noteworthy that we can reduce the total computational cost,
by employing a higher-accuracy solver and a larger timestep $\Delta t$.
Our recommendation is $(n, N) = (2,6)$ or $(4,6)$.
This is because the 4th-order and 6th-order corrections are not so much different in complexity,
and because multi cycles ($n \ge 2$) significantly improve the accuracy for large $\omega_c\Delta t$.

Owing to higher accuracy,
the hyper Boris solver allows us to use a larger timestep in kinetic simulations.
In practice, many factors would limit the timestep. 
To preserve an apparent gyroradius $r_L \approx |\vec{v}-\vec{v}_{D}|/\omega_c$,
one may want to keep $\omega_c\Delta t \lesssim 0.5$--$1.0$,
as discussed in Ref.~\cite{zeni20}. 
In PIC simulation, it would be necessary to keep $\Delta t$ small,
so that the particle either stays in the same cell or moves to the adjacent cells. 
Also, the plasma frequency $\omega_{p}$ often constrains $\Delta t$ 
%$\omega_{p}\Delta t \lesssim 0.5$
\cite{birdsall,winske96}.
Similar to the standard Boris solver,
the hyper Boris solver uses the electric and magnetic field at the particle position,
which is interpolated from the field in the grid cells. 
To take full advantage of the higher accuracy in time,
it would be useful to interpolate the electromagnetic field
at higher-order accuracy with respect to the grid size $\Delta x$.
For example, we may need a longer stencil in space \citep{lehe14}.
This also involves many other issues such as the self-force and
the charge-conservation in PIC simulation.
Coupling of the temporal and spatial accuracy needs further investigation.

%In practice, since modern PIC simulations are often memory-bounded,
%a moderate amount of computational time is acceptable,
The hyper Boris solver is nonrelativistic. 
It cannot be applied to the relativistic particle motion,
because the Lorentz factor makes the problem difficult. 
The subcycling should be useful in the relativistic regime;
however, we can no longer use the formula (Eq.~\eqref{eq:multicycle}),
because the Lorentz factor changes every subcycle.
We have also used the nonrelativistic formula (Eq.~\eqref{eq:exact})
to derive the higher-order correction.
It is not clear whether there is a simple form of a relativistic correction. 
This is left for future research. 
For a moment, we recommend the users
other Boris-type solvers \citep{umeda18,zeni18,zeni20} that
improve the Lorentz-force part of the relativistic Newton--Lorentz equation.

Finally, in addition to the Newton–Lorentz equation,
the hyper Boris solver can be applied to
the particle motion in a differentially rotating system \citep{riquelme12,hoshino13},
because the equation of motion can be transformed into
the Newton--Lorentz equation \citep[][Appendix C]{hoshino13}.
Similarly, we expect that the hyper Boris solver is useful for
differential equations in the form of $(d/dt) \vec{v} = \vec{A} + \vec{v} \times \vec{B}$.
%which is often used in spacecraft attitude \citep{hughes04,markley14}.

\section*{Author Contributions}

{\bf SZ:}
Conceptualization (lead); Formal analysis (equal); Funding acquisition (lead); Investigation (lead); Methodology (equal); Visualization; Writing – original draft; Writing – review \& editing (equal).
{\bf TK:}
Formal analysis (equal); Methodology (equal); Writing – review \& editing (equal).

\section*{Conflict of Interest}
The authors have no conflicts to disclose.

\section*{Data availability}
The data will be available from the corresponding author upon reasonable request.

\section*{Acknowledgements}
One of the authors (SZ) acknowledges S. Usami for discussion.
This work was supported by Grant-in-Aid for Scientific Research (C) 21K03627 from the Japan Society for the Promotion of Science (JSPS).

\appendix
\section{Derivation of Eq. (4)}
\label{sec:appendix}

In constant electromagnetic fields,
the equation of motion
\begin{align}
\label{eq:acc_phys2}
\frac{d\vec{v}}{d t}
&= 
\frac{q}{m} \left( \vec{E} + {\vec{v}} \times \vec{B} \right)
\end{align}
gives a gyration around the {\bf E}$\times${\bf B} velocity:
\begin{align}
\left( \vec{v}^{t+\Delta t}
 - \frac{\vec{E} \times \vec{B}}{B^2} \right)
 &= \left( \vec{v}^t - \frac{\vec{E} \times \vec{B}}{B^2} \right) \cos \theta  + \bigg( \Big( \vec{v}^t - \frac{\vec{E} \times \vec{B}}{B^2} \Big) \times \vec{\hat{b}} \bigg) \sin \theta
\nonumber\\&~~~~
+ \left( \vec{v}^t - \frac{\vec{E} \times \vec{B}}{B^2} \right)_{\parallel} (1-\cos \theta ) + \frac{q\vec{E}_{\parallel}}{m}\Delta t
.
\end{align}
where $\theta$ is given by Eq.~\eqref{eq:theta}.
We rewrite this by using $\vec{\tau}_1$ and $\vec{\varepsilon}_1$:
\begin{align}
\left( \vec{v}^{t+\Delta t}
 - \frac{\vec{\varepsilon}_1 \times \vec{\tau}_1}{{\tau}_1^2} \right)
 &= \left( \vec{v}^t - \frac{\vec{\varepsilon}_1 \times \vec{\tau}_1}{{\tau}_1^2} \right) \cos 2\tau_1  + \bigg( \Big( \vec{v}^t - \frac{\vec{\varepsilon}_1 \times \vec{\tau}_1}{{\tau}_1^2} \Big) \times \frac{\vec{\tau}_1}{\tau_1} \bigg) \sin 2\tau_1
\nonumber\\&~~~~
+ (1-\cos 2\tau_1 )
\left( \left( \vec{v}^t - \frac{\vec{\varepsilon}_1 \times \vec{\tau}_1}{{\tau}_1^2} \right) \cdot \frac{\vec{\tau}_1}{\tau_1} \right) \frac{\vec{\tau}_1}{\tau_1}  + 2 \left( \vec{\varepsilon}_1 \cdot \frac{\vec{\tau}_1}{\tau_1} \right) \frac{\vec{\tau}_1}{\tau_1}
\end{align}
This leads to the following solution.
\begin{align}
\vec{v}^{t+\Delta t}
%&=  \cos 2\tau_1 \vec{v}^t + \left( 1 - \cos 2\tau_1 \right) \frac{\vec{\varepsilon}_1 \times \vec{\tau}_1}{{\tau}_1^2}  + \bigg( \Big( \vec{v}^t - \frac{\vec{\varepsilon}_1 \times \vec{\tau}_1}{{\tau}_1^2} \Big) \times {\vec{\tau}_1} \bigg) \frac{2\sin 2\tau_1}{2\tau_1}
%\nonumber\\&~~~~
%+ \frac{(1-\cos 2\tau_1 )}{\tau_1^2}
%\left( \vec{v}^t \cdot\vec{\tau}_1 \right) \vec{\tau}_1  + \frac{ 2 \left( \vec{\varepsilon}_1 \cdot \vec{\tau}_1 \right) \vec{\tau}_1 }{\tau_1^2}
%\\
&=  \cos 2\tau_1 \vec{v}^t + 
\frac{2\sin 2\tau_1}{2\tau_1}
\bigg( \vec{v}^t \times {\vec{\tau}_1} + \vec{\varepsilon}_1 \bigg)
\nonumber\\&~~~~
+ \frac{(1-\cos 2\tau_1 )}{\tau_1^2}
\Big(
\left( \vec{v}^t \cdot\vec{\tau}_1 \right) \vec{\tau}_1 
+
\vec{\varepsilon}_1 \times \vec{\tau}_1
\Big)
+ \frac{2}{\tau_1^2}\bigg(1-\frac{\sin 2\tau_1}{2\tau_1}\bigg)
{ \left( \vec{\varepsilon}_1 \cdot \vec{\tau}_1 \right) \vec{\tau}_1 }
\end{align}
One can further rewrite this, by using the sinc function: ${\rm sinc}(x)\equiv \sin(x)/x$.
\begin{align}
\vec{v}^{t+\Delta t}
&=
\cos(2\tau_1) \vec{v}^{t}
+
2 {\rm sinc}(2\tau_1) (\vec{v}^{t} \times \vec{\tau}_1+ \vec{\varepsilon}_1 )
\nonumber \\ 
&~~~
+
2 {\rm sinc}^2(\tau_1)
\Big(
(\vec{v}^{t}  \cdot \vec{\tau}_1 ) \vec{\tau}_1 
+
\vec{\varepsilon}_1 \times \vec{\tau}_1
\Big)
+
\frac{2(1-{\rm sinc}(2\tau_1))}{\tau_1^2}
(\vec{\varepsilon} \cdot \vec{\tau}_1 ) \vec{\tau}_1 
,
\label{eq:analytic2}
\end{align}

\end{document}